\documentclass[fleqn,12pt,a4]{article}

\usepackage{fancyhdr}

\usepackage{graphicx}
\usepackage[T1]{fontenc}
\usepackage[mathbf,mathcal,text-hat-accent]{euler}
\usepackage{lmodern}
\include {epsf}

\renewcommand{\baselinestretch}{1.1}

\newcommand {\aalpha}{\mbox{\boldmath{$\alpha$}}}
\newcommand {\bbeta} {\mbox{\boldmath{$\beta$}}}
\newcommand {\zzero} {\mbox{\boldmath{$0$}}}
\newcommand {\ppp}   {{\mathcal P}}
\newcommand {\rrr}   {{\mathcal R}}
\newcommand {\aaa}   {{\mathcal A}}
\newcommand {\bbb}   {{\mathcal B}}
\newcommand {\ccc}   {{\mathcal C}}
\newcommand {\ddd}   {{\mathcal D}}

\newcommand {\xx}{\mbox{{\bf x}}}
\newcommand {\mm}{\mbox{{\bf m}}}

\newcommand {\etal}{{\it et al}, }
\newcommand {\eg}  {{\it eg} }
\newcommand {\ie}{{\it i.e.} }

\begin{document}
\title{An efficient strategy to characterize alleles and
       complex  haplotypes using DNA-markers}
\author
{
 Rodrigo Labouriau
  \thanks{Research group for Bioinformatics Genetics and Statistics,
          Department of Genetics and Biotechnology,
          Faculty of Agricultural Sciences,
          University of Aarhus.
          Corresponding author: \texttt{rodrigo.labouriau@agrsci.dk}
         }
  ,
  Poul S{\o}rensen
  \thanks{Research group for Population Genetics and Embryology,
          Department of Genetics and Biotechnology,
          Faculty of Agricultural Sciences,
          University of Aarhus.
         }
  ,
  Helle R. Juul-Madsen
  \thanks{Research group for Disease mechanisms, -markers and -prevention,
          Department of Animal Health, Welfare and Nutrition,
          Faculty of Agricultural Sciences,
          University of Aarhus.
         }
}

\date{April 2008}
\maketitle


% \newpage

\begin{abstract}
\footnotesize
\renewcommand{\baselinestretch}{0.9}
We consider the problem of detecting and estimating the strength of
association between a trait of interest and alleles or haplotypes in a 
small genomic region (\eg a gene or a gene complex), 
when no direct information on that region is available but
the values of neighbouring DNA-markers are at
hand. We argue that the effects of the non-observable haplotypes 
of the genomic regions can and should be represented by factors
representing disjoint groups of marker-alleles. 
A theoretical argument based on a hypothetical phylogenetic
tree supports this general claim. 

\renewcommand{\baselinestretch}{0.9}
The techniques described allow to identify and to infer the number of
detectable haplotypes in the genomic region that
are associated with a trait. The methods proposed use an exhaustive
combinatorial search coupled with the maximization of a version of the
likelihood function penalized for the number of parameters. 
This procedure can easily be implemented with standard statistical
methods for a moderate number of marker-alleles.
\normalsize
\end{abstract}

\renewcommand{\baselinestretch}{1.1}

\newpage

\section{Introduction}
Often in genetic applications, and in special in immune-genetics, 
interest lies in detecting and quantifying the association of a given 
genomic region to a trait. Here the genomic region might be a gene or 
a gene-complex. Typically, no bio-molecular or genomic information is directly 
available about this genomic region, but instead, a system of
DNA-markers located close to the region is used.
An example of this situation is the study presented by Schou \etal
(2007, 2008) where the possible association of the susceptibility to
several parasites and the  major histocompatibility complex (MHC) in
chickens was investigated using the microsatellite LEI0258 as a marker.
Another example involves the use of a tight
system of SNP markers to associate putative alleles in the MHC region
and susceptibility to psoriasis in humans (Orru \etal 2002).

The association between alleles or haplotypes in the genomic region of
interest and the trait is commonly characterized by a
regression-like statistical model in which the trait enters as the
dependent variable and factors representing the presence of the
marker-alleles are included among the explanatory variables.
A common practice is to declare
association between the trait and the genomic region when
at least one of the parameters representing the marker-alleles is
statistically significant. 

In order to establish association between traits and
putative  haplotypes in the genomic region of interest it is required to use a
representation of those haplotypes in terms of marker-alleles (the
only genomic information available). This
representation is crucial to properly characterize the
association. We argue that such a representation
should be constructed with groups of marker-alleles instead of only
individual marker alleles. Informally, our main point is that when considering only
groups consisted of single marker-alleles (as usually done when using a
naive approach) one might fail to represent alleles or haplotypes in
the neighbouring genomic region. This leads to a significant reduction
of efficiency or even to the complete loss of the capacity to detect
certain associations. 
Our approach requires the use of a more complex statistical inference
involving a search in a large number of possibilities. 
We show however, that the statistical
inference is feasible for a moderate number of marker-alleles (10-15 
allele-markers). 

The text is organized as follows. Section \ref{section.2} contains the
basic setup, including a description of the genetic and
molecular biological scenario, a formulation of the statistical model
in terms of a generalized linear model and some discussion on the
proper form of performing inference under those premises. A
phylogenetic based argument justifying our proposal is presented in
the last part of section \ref{section.2}.  The details
of the implementation of the statistical inference are given in
section \ref{section.3} and one examples is discussed in section
\ref{section.4}. Some discussion is provided in section \ref{section.5}. 

\section{Setup}
\label{section.2}
 
\subsection{A genetic and molecular biological scenario}
\label{subsection.2.1}

We assume that the data available consist of observations on $n$
diploid individuals from a given population for which we have the information
on the values of a trait  and a range of explanatory variables
characterizing the individuals. The interest is in detecting and
characterizing a possible association between the trait of interest
and alleles or haplotypes in a given genomic region such as a gene or
a gene complex which are not directly observable. We will refer to
these alleles or haplotypes as the {\it haplotypes in the genomic region
of interest}.

We assume additionally that data on DNA markers
located close to the genomic region is available. These markers are
assumed to be tight linked so that they can be viewed
as a single locus with several possible alleles ({\it e.g}. a
microsatellite marker or a system of very close SNP markers), called
the {\it marker-alleles}.

The data available can be thought as composed of $n$ triplets, 
\begin{eqnarray}\nonumber
  \left ( y_i, \xx_i, \mm_i \right ) , \mbox{ for } i= 1, \dots , n \, ,
\end{eqnarray}
where $i$ indexes the $n$ individuals, $y_i$ is the value of the trait, 
$\xx_i = \! \left ( x_{1i},\dots ,x_{ki} \right )$
is a vector of auxiliar explanatory variables and 
$\mm_i = \left ( m_{1i},\dots ,m_{Hi} \right )$
is a vector representing the values of $H$ allele-markers observed for the
$i^{th}$ individual.

\subsection{The basic statistical model}
\label{subsection.2.2}

We introduce below a suitable generalized linear model (GLIM) that
will serve as a framework to expose our method. It is
straightforward to extend the techniques presented to other
regression-like statistical models.

The generalized linear model describing the data is specified by
choosing a distribution for the trait among the family of the
exponential dispersion models (J{\o}rgensen \etal 1996) (typically, 
but not necessarily, 
a normal distribution) and specifying a relationship between the
expected value of the trait and the explanatory variables ($\xx$ and
$\mm$). Here we assume that there is a smooth monotone function $g$, called the 
{\it link function},
and the parameters $\bbeta = \left (\beta_1, \dots ,\beta_k \right )^T$ 
and $\aalpha = \left (\alpha_1, \dots ,\alpha_h \right )^T$ such that 
\begin{eqnarray}\label{eq.1}
 g\left [ E \left ( y_i | \xx_i , \mm_i \right ) \right ] 
\, = \,
\xx_i \bbeta +
\alpha_1 I_{m_{1i}}+ \dots + \alpha_h I_{m_{hi}}
\, ,
\end{eqnarray}
where $I_{m_{ji}}$ is an indicator variable taking the value 1 if 
the $i^{th}$ individual carries the $j^{th}$ marker-allele and 0
otherwise.
We assume, for simplicity, that all the alleles act as
completely dominant. That is, the effect of an allele in homozygous
individuals carrying two copies of the allele is equal to the effect
of the allele in heterozygous individuals carrying one copy of the
allele.
This assumption can easily be modified to include other genetic
mechanisms by modifying the definition of the factors representing 
the marker-alleles (\eg by defining factors with more than two levels
for representing partial dominance). 

Using standard techniques of generalized linear models it is possible
to make inference on the parameters $\aalpha$ and $\bbeta$. Here
our interest lies in the parameter $\aalpha$ representing the effects
of the marker-alleles, while $\bbeta$ is 
considered as an auxiliary nuisance parameter.  

The association between the genomic region of interest and the trait is often 
verified by considering a test of hypothesis given by 
\begin{eqnarray}\nonumber
  H_0: \aalpha = \zzero
  \,\,\, \times \,\,\,
  H_a: \aalpha \ne \zzero
  \,\, .
\end{eqnarray}
Since $\zzero$ represents a vector with all components equal to zero, 
the null hypothesis $H_0$ above is saying that all the
components of the vector $\aalpha$ are equal to zero while the
alternative hypothesis $H_a$ states that {\it at least one} of the marker 
alleles has a non-vanishing effect. This test can be easily carried 
out by comparing a model defined by (\ref{eq.1}) to a model defined by
\begin{eqnarray}\label{eq.2}
 g\left [ E \left ( y_i | \xx_i , \mm_i \right ) \right ] 
\, = \,
\xx_i \bbeta 
\, ,
\end{eqnarray}
using a likelihood ratio type test. Rejection of the null hypothesis
indicates association of the genomic
region in study to the trait of interest. 
Although this simple joint test detects association, it does not help
to identify the associated haplotypes in the genomic region of interest.

A naive procedure is to identify alleles or haplotypes in the genomic region by
looking for the marker-alleles with statistically
significant effects. We claim that this can be misleading. It might happen
that some of the alleles or haplotypes in the genomic region are in close linkage
disequilibrium ({\it i.e.} are tight linked to)  with  more than one
marker-allele in such a a way that in some individuals the first
marker-allele occur while the second do not occur (and vice-verse). A
phylogenetic-based argument presented below shows that this scenario can
and indeed  often occurs. Under this situation, the tests for the effect of each
single factor representing each of the marker-alleles would not have
biological meaning and would imply in a loss of
power due to a misclassification of individuals. 
Therefore, the inference on haplotypes
in the genomic region of interest should be performed using sets of marker-alleles
instead of only individual marker-alleles. 

The precise formal statement of this idea is given below.
Let $G_1, G_2, \dots , G_H$ be pairwise disjoint non-empty
subsets of the set of marker alleles $\{ m_1, \dots , m_h \}$ 
(with $H \le h$). Define the model given by
\begin{eqnarray}\label{eq.3}
 g\left [ E \left ( y_i | \xx_i , \mm_i \right ) \right ] 
\, = \,
\xx_i \bbeta +
\alpha_1 I_{G_{1i}}+ \dots + \alpha_H I_{G_{Hi}}
\, ,
\end{eqnarray}
where $I_{G_{ji}}$ is a variable taking value 1 if the $i^{th}$
individual carries at least one allele-marker belonging to the 
subset $G_j$ (for $j=1, \dots , H$).
Clearly, the simple model given
by (\ref{eq.1}) is contained in the class of models in the form given
by (\ref{eq.3}), since the disjoint subsets $G_1, G_2, \dots , G_H$
can be all constituted of a single element. However, this class of
models contains many other models (any possible combination of
non-empty disjoint subsets of $G$), which opens the possibility of
finding a model of this type that suitably describes the genetic
phenomena in play. We discuss in section \ref{section.3} a strategy
for searching for the best representation among the many
possibilities. 

\subsection{A phylogenetic-based argument}
\label{subsection.2.4}

A number of special structures naturally appear during 
the evolution process of a population. As a consequence, the
information that DNA-markers carry on neighbour loci is distributed 
according to some characteristic patterns. In this section we illustrate this 
general claim using a simple phylogenetic-like construction based on dichotomous
branching trees. We will show how some motifs of association involving
markers and alleles in the genomic region occur naturally. This
will then be used to argue in favor of using a
proper representation of the effect of DNA-markers and to base the 
inference using statistical models defined with groups of marker-alleles as in
(\ref{eq.3}).

Consider a locus $\rrr$ in the genomic region of interest and two
observable markers $\aaa$ and $\bbb$ in a neighbourhood of
$\rrr$. Suppose, for exposition simplicity that $\rrr$, $\aaa$ and
$\bbb$ are di-allelic with pairs of alleles $(R,r)$, $(A,a)$ and
$(B,b)$ respectively. Assume, moreover, that recombination between these
loci can be neglected due to a strong linkage disequilibrium around the
region of interest. We can think of each of those alleles as the result of
a single event occurred at some point in the evolutionary history of
the population in play ({\it e.g.} a single nucleotide mutation or the
duplication of a small genomic region). The sequence of events that
generated these alleles can be represented by a tree with three
dichotomous branching, each branching corresponding to the event that generated
one of the alleles. We use the convention that the alleles
represented by capital letters are the results of events, while the
alleles represented by small letters are the reference alleles,
or wild types, corresponding to the states of the loci before the
events.

A marker-allele $A$ {\it carries information} about the locus $\rrr$
when the knowledge of the occurrence of $A$ determines the allelic
form of $\rrr$. If the allele $A$ can occur together with the allele
$R$ and the allele $r$, then $A$ is said to be {\it neutral} with respect to the
locus $\rrr$.  For instance, if the branching that formed the locus
$\rrr$ occurred before the branching of the locus $\aaa$ and the
branching of $\aaa$ occurred in the branch of the tree containing the
allele $R$ (see Figure 1A), then the occurrence of the allele $A$ in
the locus $\aaa$ implies that the locus $\rrr$ carries the allele
$R$. Moreover, in that circumstances the occurrence of the allele $a$
does not imply neither that $\rrr$ carries the allele $R$ nor
$r$. Therefore the allele $A$ carries information on the locus $\rrr$
and the allele $a$ is neutral with respect to $\rrr$.

Figure 2A illustrates a scenario where the branching of the locus
$\rrr$ occurred first, which was followed by the branching of the
locus $\aaa$ and then the branching of the locus $\bbb$. Moreover, the
branching of the locus $\aaa$ occurred in the branch of the tree containing
the allele $R$ while the branching of the locus $\bbb$ occurred in the
branch of the tree containing the marker-allele $a$. Under these
circumstances there are only four possible haplotypes: $AbR$, $aBR$,
$abR$, $abr$. The allele $A$ only occurs together with the
allele $R$, and it carries information about the locus
$\rrr$. Analogously, the allele $B$ carries also information on the locus
$\rrr$. Since the alleles $a$ and $b$ might occur together with both the
allele $R$ and the allele $r$, then both $a$ and $b$ are neutral with
respect to the locus $\rrr$. We can still draw further conclusions
about the distribution of the information on the locus $\rrr$. If we
want to use a rule for detecting the presence of the allele $R$ based
on of the occurrence of marker-alleles, then the rule '$R$ occurs
when $A$ or $B$ occurs' would detect two out of the three occurrences
of the allele $R$. A rule based only on the occurrence of the
marker-allele $A$ would only detect one out of the three possible
occurrences of the allele $R$ and therefore would be less
efficient in detecting $R$ than the rule using the alleles $A$ and
$B$ together. The alleles $a$ and $b$ are both neutral and therefore the
occurrence of the allele $R$ in the haplotype $abR$ cannot be detected
using the information contained in the marker-alleles. We conclude
that under the scenario described by Figure 2A, one can only detect
the occurrence of the allele $R$ using a rule based on the
marker-alleles in two out of the three possible haplotypes containing
$R$. This maximum possible information recovery is attained only by
the rule '$R$ occurs when $A$ or $B$ occurs'.

A different scenario is described in Figure 2B where the branching in
the locus $\aaa$ occurred first, followed by the branching in the locus
$\rrr$, in the branch containing the allele $A$, and then the
branching in the locus $\bbb$ in the branch of $\rrr$ containing the
allele $R$. In this case the four possible haplotypes are: 
$ABR$, $AbR$, $Abr$ and $abr$. Therefore the marker-allele $B$ carries information
on $\rrr$ and the marker-allele $A$ is neutral. The alleles $a$ and $b$ necessarily
occur together and both carry information on $\rrr$ (but the same
information). Here there are two cases in which the genotype of the
locus $\rrr$ is determined by the genotypes of the marker-alleles:
'occurrence of $B$' and 'occurrence of $a$ and $b$' implying the
occurrence of $R$ or $r$ respectively. Note that the last event
'occurrence of $a$ and $b$' is equivalent to the event
'not occurrence of $A$ or $B$'. Under this scenario there are only two
rules based on the marker-alleles genotypes that determine the genotype at
the locus $\rrr$, both can be expressed as the effect of a combination
of the occurrence of the marker-alleles $A$ and $B$. The first rule
('occurrence of $B$ implies occurrence of $R$') can also be expressed
as the effect of a single allele-marker as in the traditional
inference method, but the second rule ('not occurrence of $A$ or $B$
implies the occurrence of $r$') requires the use of groups of
marker-alleles as in the models described in (\ref{eq.3}) to be
properly represented in a
statistical model. Here sticking only to rules based on single markers
would represent a loss of half of the possibilities for determining the
allele at the locus $\rrr$, that is a loss of half of the information
on the genotype of the locus $\rrr$ that could be recovered with the
knowledge of the marker-alleles.

Figure 3 displays the branching tree of a more complex scenario
composed with the locus $\rrr$ in the genomic region of interest and
four marker-loci $\aaa$, $\bbb$, $\ccc$, and $\ddd$, with alleles $A$,
$a$, $B$, $b$, $C$, $c$, $D$ and $d$ respectively (we apply the same
notational convention as before). The following six haplotypes are
formed: $AbcdR$, $aBCdR$, $aBcdR$, $abcDR$, $abcdR$ and
$abcdr$. Therefore, the rule 'if $A$ or $B$ or $C$ or $D$ occur then
$R$ occurs' detects four out of the five haplotypes containing the
allele $R$. Moreover, under the current scenario, this is the best
possible rule that could be constructed with the information on the
marker-alleles that we have at hand. Although the allele $C$ carries
information on the locus $\rrr$, this information is redundant since
$C$ occurs always together with $B$. We can then remove the occurrence
of the allele $C$ from the rule and still detect the same cases
where the allele $R$ occurs when using the rule including the 
marker-alleles of the four marker loci.

The discussion above shows that in several situations the use of a
naive representation of the effects of the marker-alleles is
inefficient and that fully efficiency is obtained when using the
approach involving the representation of groups of
allele-markers. This phenomenon is not restricted to the few
scenarios presented here. We claim, without giving a formal proof, that
every time that there is a branching after the branching that
generated the allele in the locus $\rrr$ the new marker-allele formed
will carry information on the locus $\rrr$. Moreover, if the branching
occurs in the branch that contains the wild type allele of the last
branching of a marker locus, then the new marker allele formed will
add new information on the locus $\rrr$ that is not contained in the
marker-alleles formed before. This progressive gain of information
obtained while the new marker-alleles are being formed is only fully
realized if we use a rule of the type 'if $A$ or $B$ or $C$ or
... occurs, then $R$ occurs'. 

\vspace{5cm}

% ---------------
% \newpage
\begin{figure}[h!]
  \centering
   \includegraphics[width=6cm]{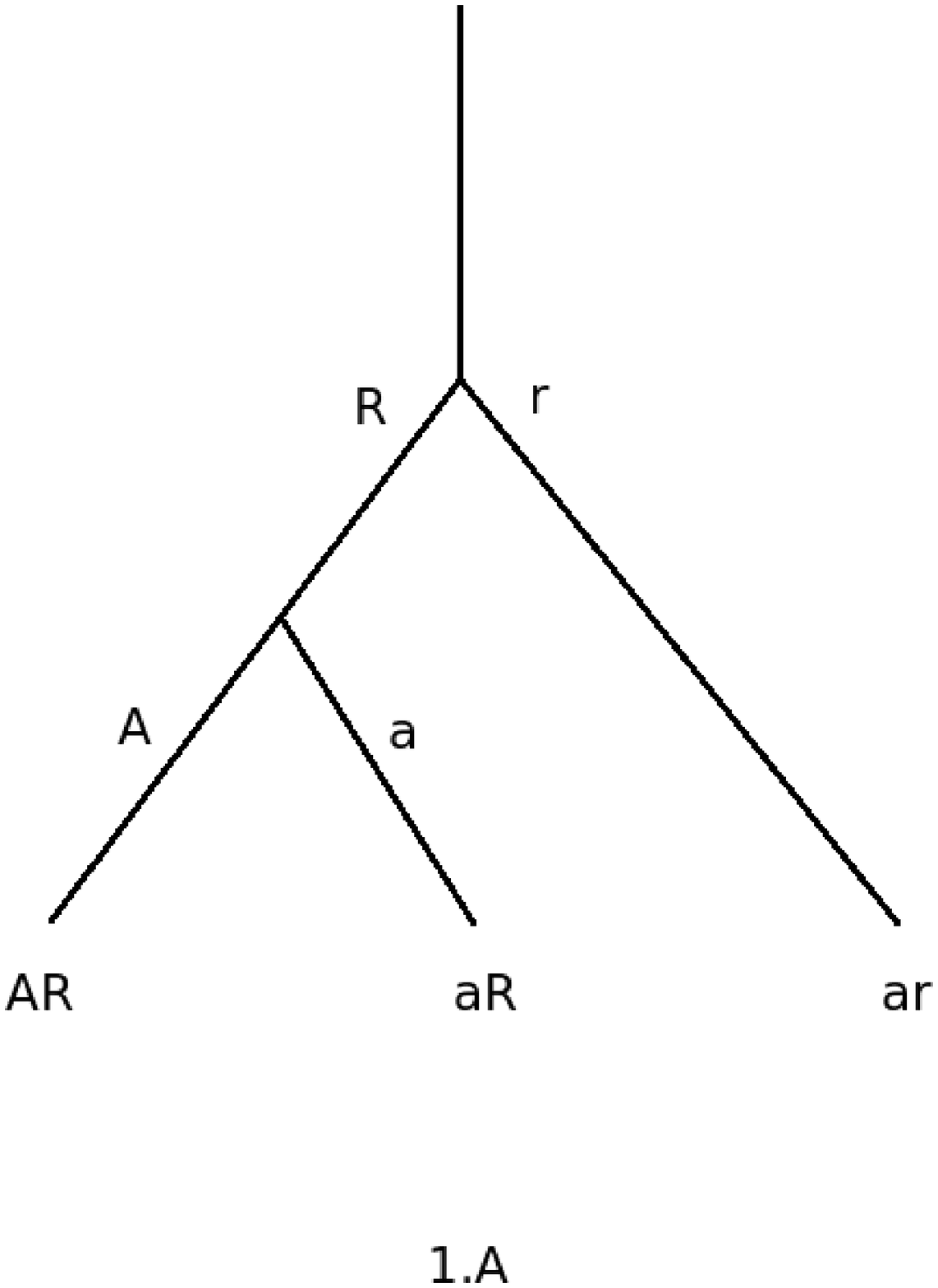}
   \includegraphics[width=6cm]{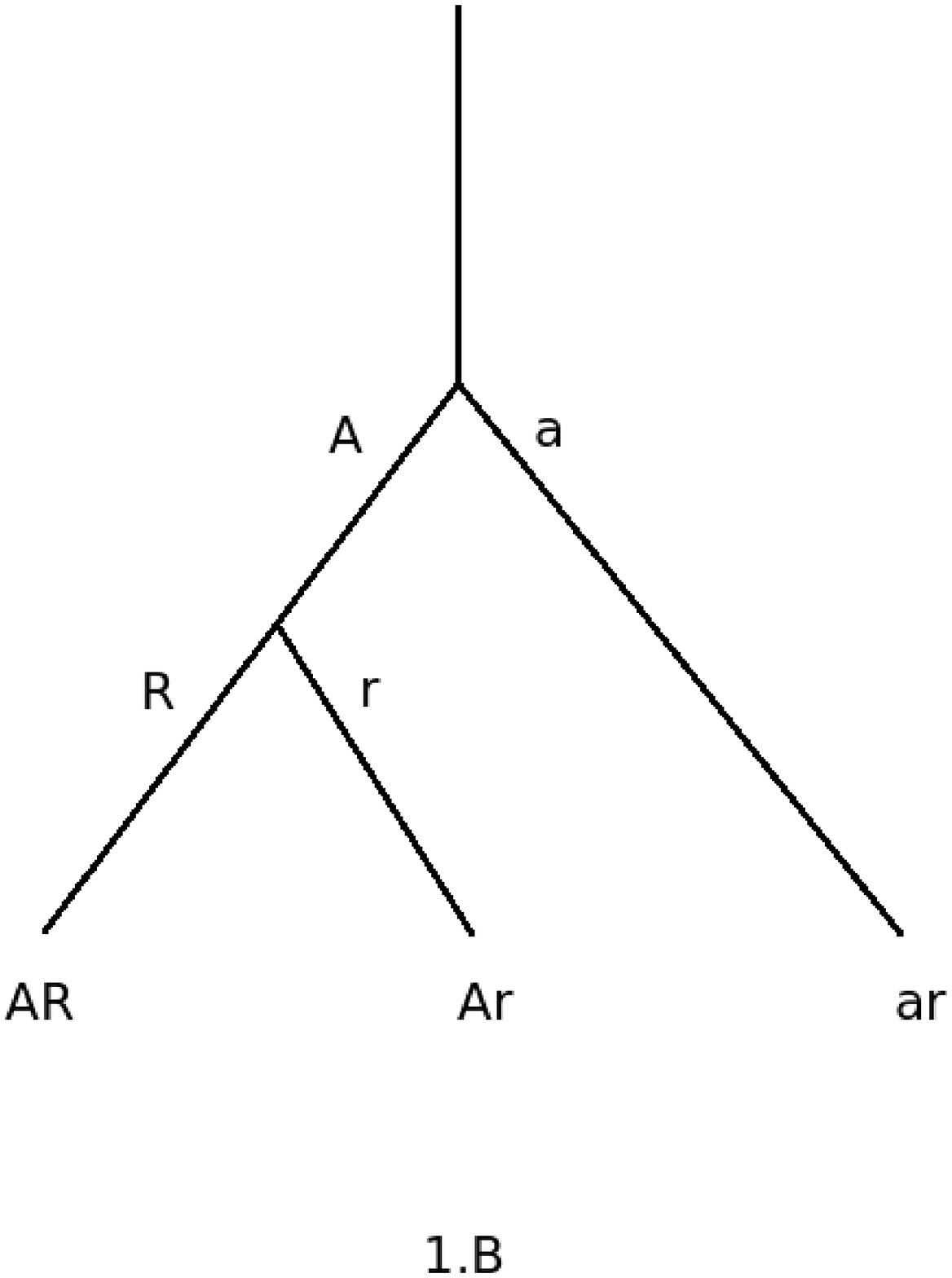}
  \caption{\footnotesize
  Two dichotomic phylogenetic trees involving a locus $\rrr$ in the
  genomic region of interest with the alleles $R$ (variant allele) and
  $r$ (wild type allele) and one marker loci $\aaa$ with
  alleles $A$ and $a$. The haplotypes formed at
  each ending branch are displayed at the bottom of the tree.}
   \label{Figura.1}
\end{figure}
% ---------------

% \clearpage

% ---------------
% \newpage
\begin{figure}[h!]
  \centering
   \includegraphics[width=6cm]{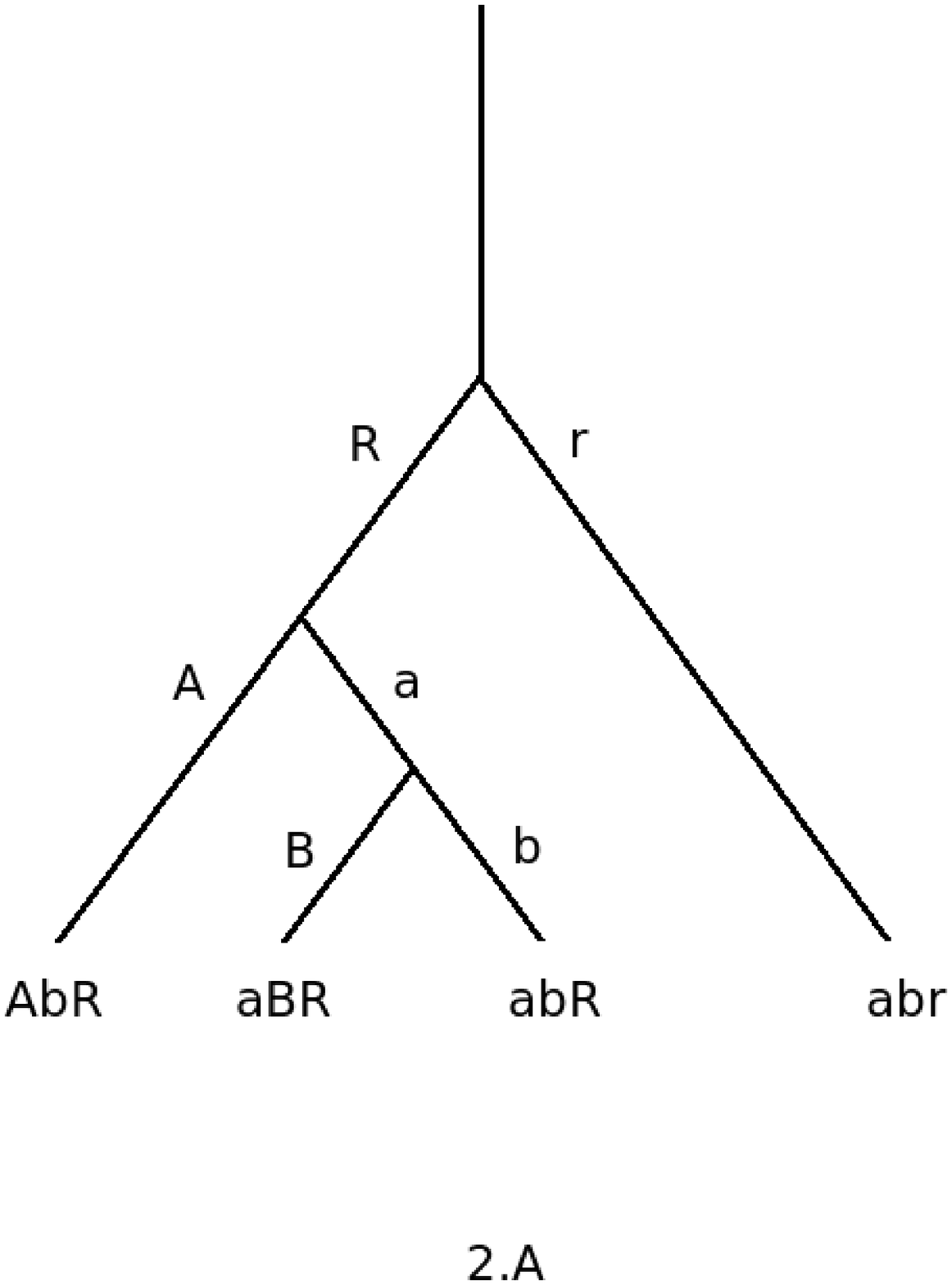}
   \includegraphics[width=6cm]{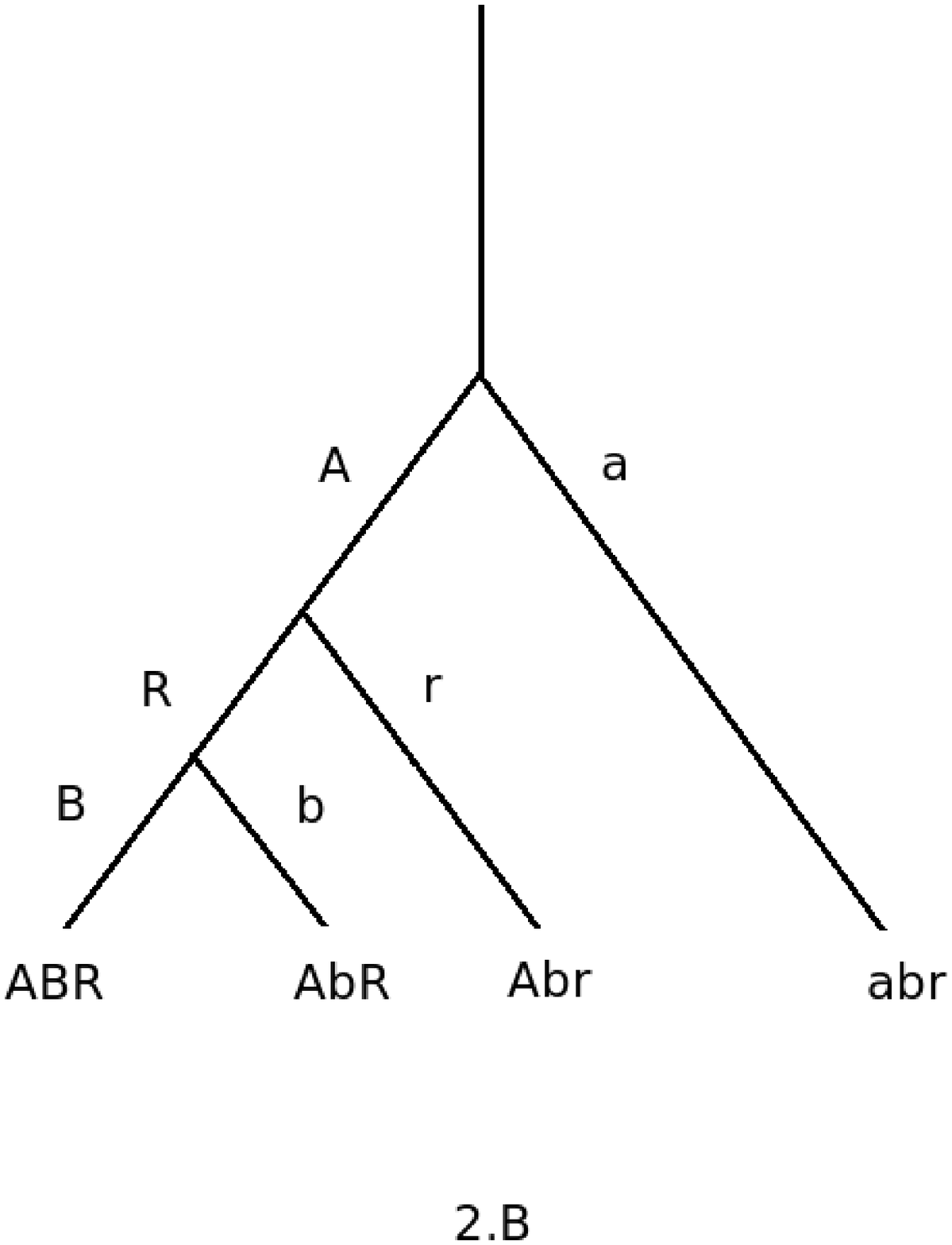}
   \includegraphics[width=6cm]{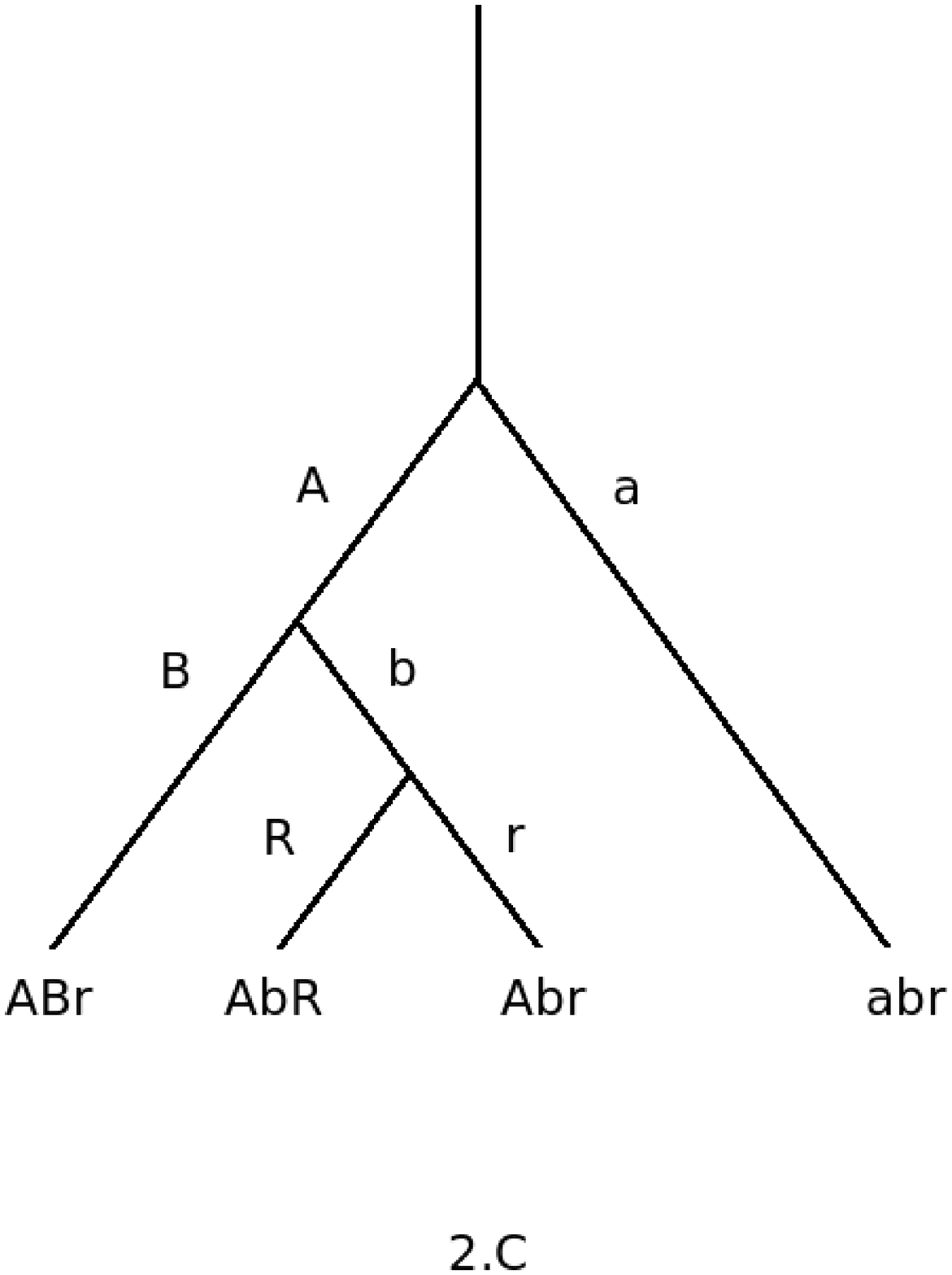}
  \caption{\footnotesize
  Three dichotomic phylogenetic trees involving a locus $\rrr$ in the
  genomic region of interest with the alleles $R$ (variant allele) and
  $r$ (wild type allele) and two marker loci $\aaa$ and $\bbb$ with
  alleles $A$, $a$, $B$, $b$ respectively. The haplotypes formed at
  each ending branch are displayed at the bottom of the tree.}
   \label{Figura.2}
\end{figure}
% ---------------

% ---------------
% \newpage
\begin{figure}
  \centering
   \includegraphics[width=9cm]{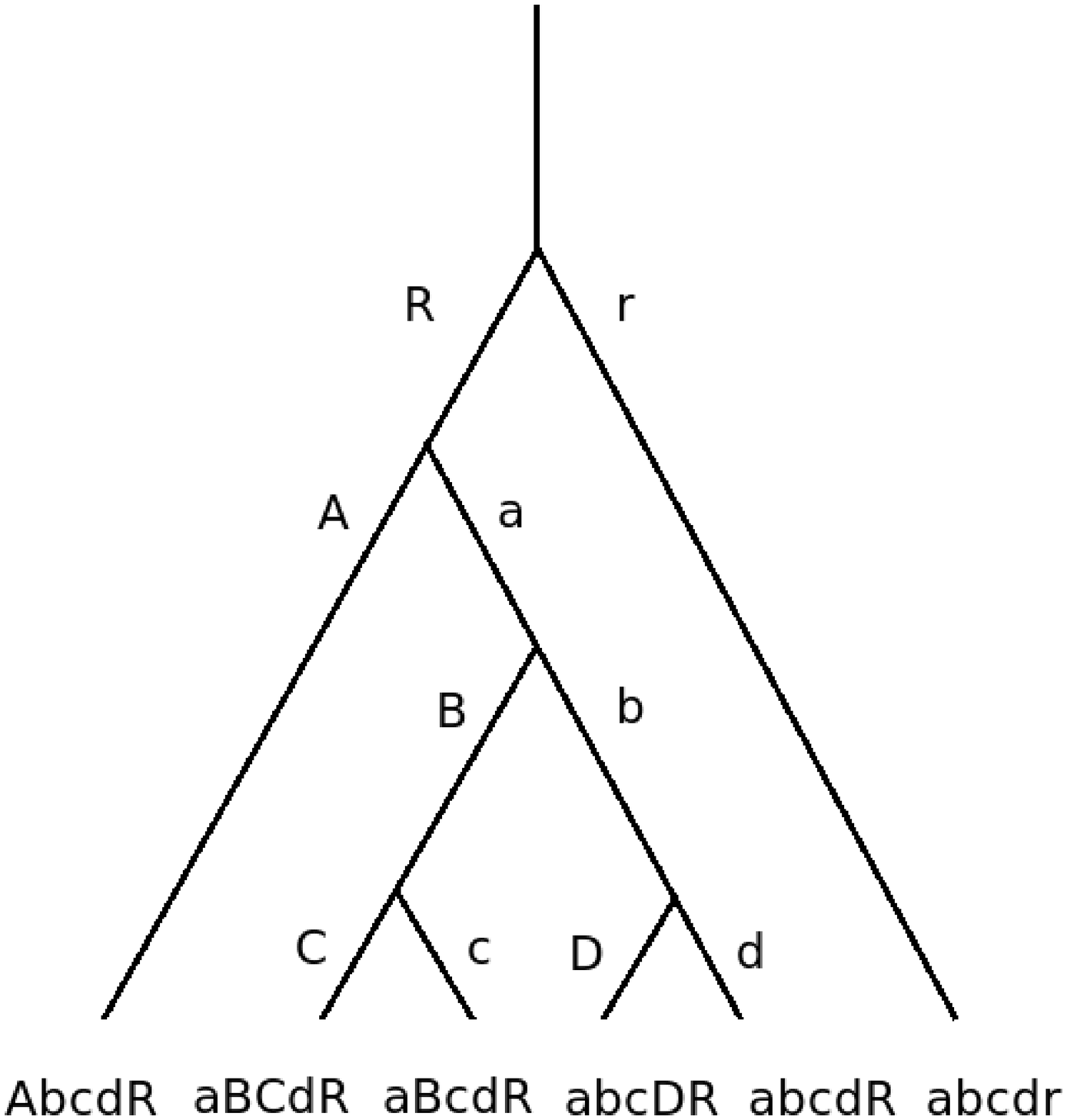}
  \caption{\footnotesize
  A complex dichotomic phylogenetic tree, involving a locus $\rrr$ in the
  genomic region of interest with the alleles $R$ (variant allele) and
  $r$ (wild type allele) and four marker loci $\aaa$, $\bbb$, $\ccc$
  and $\ddd$. The haplotypes formed at
  each ending branch are displayed at the bottom of the tree.}
   \label{Figura.3}
\end{figure}
% ---------------

\clearpage

\section{Inference with a moderate number of  \\
marker-alleles}
\label{section.3}

\subsection{Exhaustive search strategy}
\label{subsection.3.1}

The strategy we propose for characterizing the association between a
genomic region and a trait consists in searching exhaustively all the
possible groupings formed with sets of marker-alleles and then choose
the best candidate among the many possibilities. Here the best
candidate is one that represents all the haplotypes of the genomic
region of interest that are associated with the trait and that is not
redundant.
We define a {\it grouping of the marker-alleles} as a collection of
non-empty disjoint subsets of
the set of all marker-alleles. The subsets of a grouping are called 
{\it  marker-alleles groups} (MAG). The idea is to use MAGs to
represent haplotypes in the genomic region of
interest that might be associated with the trait. For each possible
grouping of the marker-alleles one statistical model of the type
described by (\ref{eq.3}) containing factors representing each MAG of
this grouping is fit. The grouping that generates the model with the
best fit, according to a criterion to be defined below, is chosen to 
represent the association between the genomic region of interest and
the trait. The grouping will be chosen in such a way that it will not
contain redundancy, so each MAG will represent one haplotype in the
region of interest. The number of MAGs in this grouping will be the
number of detectable haplotypes associated with the trait. The
magnitude of the effect of each MAG will be then the component of the
magnitude of the haplotype that is detectable through the
marker-alleles (which is smaller or equal to the magnitude of the
effect of the unknown haplotype).
This procedure is only feasible
if the number of marker-alleles is not very large (we were able to
analyse a data with 9 allele markers in few minutes in an ordinary
personal computer).

It is convenient to make the exhaustive enumeration of all possible
groupings of the marker-alleles in the following way.
Consider the class $\ppp$ of all subsets of the set of
marker-alleles $G = \{ m_1 , \dots , m_h \}$. 
Formally, $\ppp$ is the class of parts of $G$.
We associate one statistical
model to each element $c$ of $\ppp$ by defining the model of the type
defined by (\ref{eq.3}) that incorporate
factors representing the groups of marker-alleles present in
$c$. For $j = 1, \dots , h$ let
$\ppp_j$ be the class of all the subsets of $G$
containing exactly $j$ disjoint non-empty sub-sets. Clearly $\ppp$ is
the disjoint union 
\begin{eqnarray}\label{eq.4}
  \ppp = \ppp_1 \cup \dots \cup \ppp_h \, .
\end{eqnarray}
Therefore we can search for the best models by proceeding in two
steps: First we find the best model for each $\ppp_j$ ($j = 1, \dots,
h$) and then we find the best model among the candidates found in the
first step. The selection of the best model related to a given $\ppp_j$ ($j = 1, \dots,
h$) is done by choosing the model with the largest value of the
likelihood (or equivalently the log-likelihood)
function. In this way the set of values of the likelihoods of the
chosen candidate for each $\ppp_j$ is a profile set and plays a rule
analogous to the rule of a profile likelihood curve for the number of
marker-alleles. Denote the model that attains the maximum
of the likelihood for a given  $\ppp_j$ by $M_j$ and the value
log-likelihood function of $M_j$  at the maximum by $\hat l_j$ (for $j
= 1, \dots,h$).  We refer to these quantities as the  {\it
  profile model} and the {\it profile log-likelihood} of order $j$.

\subsection{Determination of the number of detectable associated haplotypes}
\label{subsection.3.2 }

The next step in the procedure of inference is to choose the class
$\ppp_j$ ($j = 1, \dots,h$) that yields the best statistical model.
If we assume that the haplotypes in the genomic 
region of interest are representable in terms of subsets of  $G$, then choosing the
class $\ppp_j$ that produces the best statistical model is equivalent to
infer the number of detectable haplotypes in the genomic
region of interest.

The profile log-likelihood
never decreases when the number of haplotypes assumed in the model increase, {\it i.e.}
\begin{eqnarray}\label{eq.5}
 \hat l_1 \le \dots \le \hat l_h
 \,\, ,
\end{eqnarray}
since a model $M_i$ ($i <h$) can be expressed as a sub-model of a
model with $i+1$ MAGs in which a pair of MAGs present the same effect.
As a consequence, it is not reasonable to estimate the number of
haplotypes in the region of interest by choosing the 
$\ppp_j$ with larger log-likelihood. We argue next that maximizing the
negative Akaike information (or a variant of it) is
a reasonable procedure for inferring the number of haplotypes
in the genomic region of interest.

The inequality (\ref{eq.5}) does not extract all the information
available on the development of the
profile likelihood curve as the number of putative haplotypes
of the genomic region of interest increases. Indeed, the profile
log-likelihood curve is expected to increase significantly with the
number of putative haplotypes until the number of detectable
haplotypes is reached. After that point the  profile
log-likelihood curve is supposed to remain approximately constant. To see that,
consider the situation where there are $J$ associated haplotypes in
the genomic region. If $j < J$, then the model $M_j$ fails to
represent at least one haplotype and then the profile log-likelihood
should increase in a statistically significant way with the addition
of the possibility to represent one more haplotype. Once attained the
number of haplotypes the gain obtained by increasing the capacity of
the model to represent one more haplotype vanishes and only marginal
gains in the profile log-likelihood are expected (see figure 4). We assume implicitly
here that there
are no significant mixtures in the data and that the model is not
missing any important explanatory variable (see Figure 4)
The informal argument above suggests we can infer the number of
detectable haplotypes in the genomic region by searching for
the point at which the profile log-likelihood curve remains
(approximately) constant. One way to do that is to subtract a suitable
quantity from the profile log-likelihood. By doing that, the new
adjusted profile log-likelihood would decrease approximately linearly 
in the region where the original profile log-likelihood was
constant (\ie when the assumed number of
haplotypes is larger than the number of detectable
haplotypes in the genomic region of interest). If the
subtracted quantity is not too large,
the adjusted profile log-likelihood curve will still increase in the
region where the original profile log-likelihood is increasing
significantly (before attaining the number of detectable
haplotypes). The so called Akaike information criterion
(Akaike, 1974, Burnham and Anderson, 2002) explores this idea and is
equivalent to subtract the number of parameters in the model from the
log-likelihood. In fact the Akaike information $AIC$ is defined as
minus twice the difference of the log-likelihood and the number of
parameters, more precisely,
\begin{eqnarray}
  \nonumber
  AIC = -2 \left \{ \log ({\hat L}) - k \right \} 
  \, \, ,
\end{eqnarray}
where $k$ is the number of parameters in the statistical model, and
${\hat L}$  is the maximized value of the likelihood function for the estimated model.
Minimizing the Akaike information is equivalent to maximizing the
log-likelihood adjusted by subtracting the number of parameters in the
model. This apparently arbitrary choice for the quantity subtracted from the
log-likelihood can be justified as being equivalent to subtract
from a likelihood ratio statistic its (asymptotic) expected value. 
Alternatively, one might subtract $3.85 k$ from the profile log-likelihood which would be
equivalent to perform a likelihood ratio test for incorporating the
representation of an additional haplotype to the current model
when working at a significance level of 5\%. 

Summing up, the procedure we propose is to maximize the log-likelihood
in each class $\ppp_i$ (which is equivalent to minimize the AIC in
this class of models since all the models in $\ppp_i$ has by
construction the same number of parameters), for $i=1,\dots ,h$, and
then choose the model
with the smaller (profile) AIC (or equivalently, with the larger adjusted
profile log-likelihood). 

\section{Example: Chicken susceptibility to \\ helminths}
\label{section.4}

The association between the susceptibility to the helminth {\it
Ascaridia galli} in chickens and the major histocompatibility (gene)
complex (MHC) was investigated in two recent studies (Schou \etal
2007, 2008). These studies used the microsatellite
LEI0258 (Fulton et al., 2006), which is located in a non-coding region between two
contiguous regions of the MHC gene complex (the B-F/B-L and the BG loci),
to obtain eight polymorphic marker-alleles here denoted
195bp, 207bp, 219bp, 251bp, 264bp, 276bp, 297bp and 324bp. 
Since recombination within the chicken MHC is very rare (Plachy
et al., 1992; Miller et al., 2004), the alleles of the microsatellite
LEI0258 are expected to be in tight linkage disequilibrium with haplotypes formed
by alleles at the BF/BL and the BG loci (\ie the MHC
haplotypes). Moreover, we can discard the possibility of a direct
effect of the LEI0258 alleles since this marker is located in a non-coding
region (as any microsatellite marker). Therefore it is reasonable to
apply the techniques described
above using the alleles of the microsatellite LEI0258 as
marker-alleles in the set-up described above.

In the first study (Schou \etal 2007) the intensity of infection with 
{\it A. galli} was determined for birds of two chicken
breeds by counting the number of this worms  found in the
intestine of each of the birds examined. The counts were categorized as,
zero, low (up to 3 counts), intermediate (4 to 10 counts) and high
(more than 10 counts). The cut-off points used for defining the
categories above were chosen in such a way that the losses of the
K\"{u}lback-Leiber information about the counts due to a
discretization were minimized.
The association between the intensity of infection
and the MHC haplotypes was studied by applying
a baseline-category logits model for multinomial
distributed data (Agresti, 1990) using the zero-category as a
reference.  Inference in these models can be performed by
fitting three logistic models constructed with a common reference
category (Agresti, 1990) which can be done by using standard
generalized linear models (with a binomial distribution and a
logistic link). The standard method was used in this study
and an association was declared if the effect of a marker-allele, in
the presence of the other marker-alleles, was statistically
significant. Using this procedure it was found that the occurrence of
the marker-allele 276bp was associated with an increased
resistance. No further associations were found with the standard
procedure. 

\begin{figure}
  \centering
   \includegraphics[width=12cm]{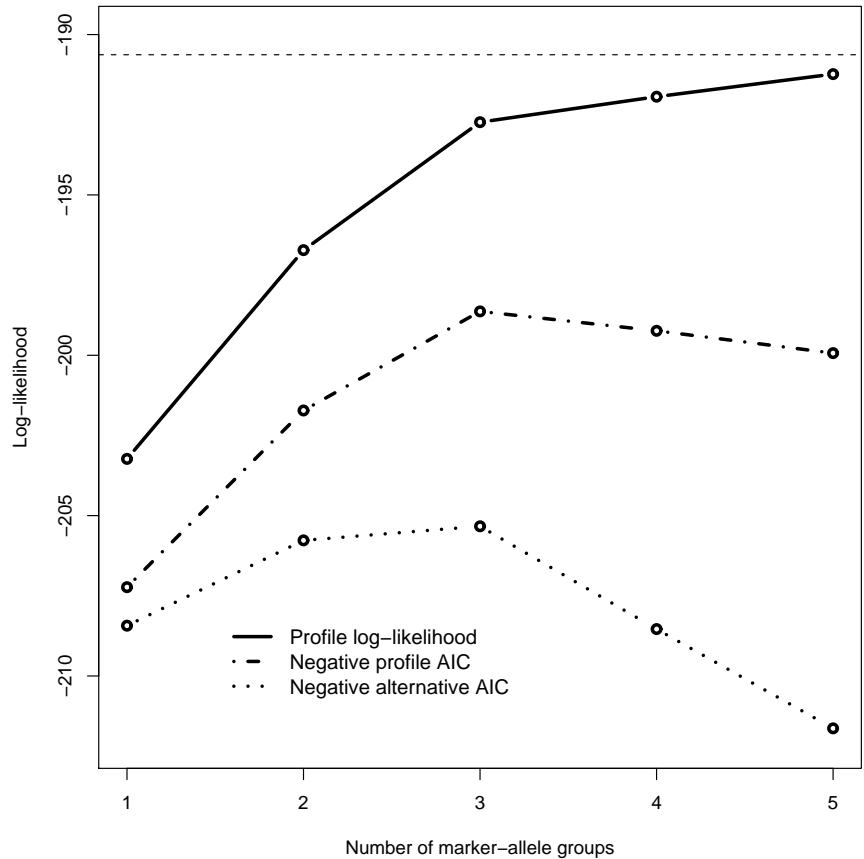}
  \caption{\footnotesize
  Profile log-likelihood, negative profile Akaike information
  (dashed and dotted line) and negative profile alternative Akaike
  information (dotted line) as
  a function of the assumed number of marker-allele groups for
  the data of the second study on susceptibility to {\it A. galli}
  (Schou \etal 2008).
  The dashed horizontal line represents the log-likelihood
  of a model containing the maximum number of marker-allele groups
  (\ie 8) evaluated at its maximum, that is, an upper bound for the
  values of the profile log-likelihood. The alternative Akaike
  information was obtained by subtracting $3.85 k$ for each successive increase in the 
  number of marker-allele groups, corresponding to test the
  incorporation of one more marker-allele group at a 5\% level of
  significance. The value 10 was added all the values of the negative
  alternative profile Akaike information for graphical
  convenience. The negative profile Akaike information and the
  alternative negative profile Akaike information presented both a
  maximum at 3 suggesting  the presence of 3 detectable marker-allele
  groups.
}
   \label{Figura.4}
\end{figure}

A second study (Schou \etal 2008) was independently performed with
different birds of the same two chicken breeds.
In this study the birds were
inoculated with {\it A. galli} under controlled experimental
conditions and the fecal excretion of {\it A. galli} eggs was
monitored. Each animal was classified based on the counts of eggs as
presenting zero, low,
intermediate or high infection level.  A baseline-category logits
model was applied but, differently from the first study, using the
strategy of searching for  marker-allele groups (MAG).  Three
marker-allele groups, denoted MAG-1, MAG-2 and MAG-3, were
identified and found to be associated with the intensity of infection
with {\it A. galli}. Figure \ref{Figura.4} displays the profile log-likelihood and
profile Akaike information as a function of the number of MAGs assumed
to be associated to the infection level. A joint
likelihood ratio test indicated a statistically significant effect of
these three MAGs on the intensity
of infection (p=0.0013). MAG-1 was formed by the LEI0258 alleles 297bp
and 324bp; MAG-2 was composed of the alleles 195bp, 207bp, 219bp and
264bp; and MAG-3 only contained
the allele 276bp. Detailed analyses revealed that animals carrying MAG-1
or MAG-3 presented larger resistance to {\it A. galli}, while MAG-2 was
associated with augmented susceptibility (Schou \etal 2008).

An {\it a posteriori} analysis of the data of the first study using
the strategy of searching for marker-allele groups yielded the same
significant marker-allele groups with MAG-1 and MAG-3 associated to
resistance and MAG-2 associated to susceptibility to {\it
  A. galli} (reported in Schou \etal 2008). Note that when applying
the standard strategy only the effect of the marker-allele 276bp was
found significant, which is equivalent of detecting the marker-allele
group MAG-3 (composed only by the allele 276bp). None of the
marker-alleles composing the marker-allele group MAG-1 and MAG-2 (\ie
297bp, 324bp, 195bp, 207bp, 219bp and 264bp) presented an individual
significant effect, which illustrates the loss of power to detect
association  when applying the standard modelling strategy.

% ---------------
% \newpage
% ---------------
\clearpage

\section{Discussion}
\label{section.5}

We presented a strategy for performing statistical inference that
allows to represent the occurrence of non-observable alleles or
haplotypes in a genomic region of interest in terms of a range of observable
marker-alleles at highly linked loci. The kernel idea presented here is that the natural
unit to build statistical models under this context are not the
marker-alleles but groups of marker-alleles. As argued 
due to the way haplotypes in a genomic region of
relatively small size (such that allow us to ignore recombination) are
formed during the evolution of a population, certain haplotypes formed
with marker-alleles will occur naturally in tight linkage
disequilibrium with (non-observable) haplotypes in the genomic
region. The way these marker-allele haplotypes are constituted imply
that detection rules based on indication functions of groups of
marker-alleles are optimal in the sense that they allow to extract the
maximum possible amount of information that is contained in the
marker-alleles. Naive representations constructed exclusively with
groups of allele-markers with only one element are bounded to use inefficiently the
information (if not destroy it completely), as illustrated in an
example with real data.

The techniques presented here involve fitting many models
and selecting a best candidate among the (very) many
possibilities, following a sequence of models with increasing number
of assumed marker-allele groups. This order is tough to facilitate
the inference of the number of marker-allele groups with  {\it
  detectable} effect on the trait of interest.
However, this {\it force brut} exhaustive search is not
feasible for a large number of marker-alleles, since the number of
possibilities to be checked increases very rapidly with the number of
marker-alleles. An alternative is to use Monte Carlo based algorithms
for maximization in discrete parameter space as simulated annealing or
the genetic algorithm.

We showed that the classical criterion of maximizing the
log-likelihood cannot be used to estimate the number of detectable
marker-alleles groups, since the likelihood function cannot decrease
with the number of MAGs. A way around that is to use the Akaike
information criterion which penalizes the log-likelihood of models
with too many (unnecessary) parameters. Other forms of penalized
likelihood could be applied, as for instance the alternative
information criterion proposed in the text, which is equivalent to
perform a likelihood ratio test for incorporating one extra MAG in the
model (at a 5\% level of significance). The choice of the penalized
likelihood method to be used depend on the type of basic statistical
model used.

We used generalized linear models to explain our ideas, but we
stress that other models could had been used without essentially
changing the procedure exposed here. Indeed, the example presented uses
in fact a slight extension of generalized linear models. Probably one
of the most relevant extensions would be the
incorporation of random components allowing to represent 
population structure, co-ancestry and polygenic effects. Another
possibility to be explored is the incorporation of information on
ancestors genotypes and other mechanisms of inheritance. In
conclusion, the techniques presented here are flexible and relatively
easy to implement using classical statistical models and standard
software.
 
\vspace{1cm}

\noindent
\Large {\bf Acknowledgements:} \normalsize \\
This work originated during the statistical analysis of the data
collected by Torben Wilde Schou. We thank him for kindly allowing us to
present part of his data as an example.

\end{document}